# Two-Dimensional graphene-HfS$_2$ van der Waals heterostructure as electrode material for Alkali-ion batteries

Gladys King'ori[a, b], Cecil N M Ouma[c], Abshiek Mishra[d], George O Amolo[b], Nicholas Makau[a].

[a]University of Eldoret, P.O Box. 1125 – 30100, Eldoret, Kenya.
[b]Technical University of Kenya, Haile Selassie Avenue, P.O Box 52428 - 00200, Nairobi, Kenya.
[c]HySA-Infrastructure, North-West University, Faculty of Engineering, Private Bag X6001, Potchefstroom, South Africa, 2520
[d]University of Petroleum and Energy sciences, Physics Department, Energy Acres, Bidholi via Premnagar, Dehradun 248007, UK, India.

**Abstract**

Poor electrical conductivity and large volume expansion during repeated charge and discharge is what has characterized many battery electrode materials in current use. This has led to 2D materials, specifically multi-layered 2D systems, being considered as alternatives. Among these 2D multi-layered systems are the graphene-based van der Waals heterostructures with transition metal di-chalcogenides (TMDCs) as one of the layers. Thus in this study, graphene-Hafnium disulphide (Gr-HfS$_2$) system, has been investigated as a prototype Gr-TMDC system for application as battery electrode. Density functional theory calculations indicate that Gr-HfS$_2$ van der Waals heterostructure formation is energetically favoured. In order to probe its battery electrode applications capability, Li, Na and K intercalants were introduced between the layers of the heterostructure. Li and K were found to be good intercalants as they had low diffusion barriers as well as positive open circuit voltage. A comparison to bilayer graphene and bilayer HfS$_2$ indicate that Gr-HfS$_2$ is a favourable battery electrode system.

## 1.0 Introduction

Rechargeable battery electrode materials suffer from poor electrical conductivity and large volume expansion during repeated charge and discharge, which neutralizes their large capacity and impairs the long term electrochemical stability [1]. This has led to studies on how electrode materials can be modified either via doping or creation of Gr based two-dimensional (2D) van der Waals heterostructures, notably those based on transition metal di-chalcogenides (TMDCs). 2D van der Waals heterostructures affords an opportunity to develop rechargeable battery storage systems with high rate capacity and storage density as well as cyclic stability [2][3]. Due to the challenges facing electrode materials such as low gravimetric and volumetric energy densities, there is need for materials with possible higher gravimetric and volumetric energy densities. However, many of them suffer from; limited electrical conductivity, slow lithium transport, large volume expansion, low thermal stability, mechanical brittleness, dissolution as well as other unsuitable interactions with the battery electrolyte [4].

2D materials offer several favorable properties over their 3D counterparts especially in the design of next generation devices [5]. Graphene a pioneer 2D material has been widely investigated due to it being very thin, highly transparent, very flexible, having large surface area, outstanding conductivity [6] and good stability for chemical agents [7]. These properties make it suitable for transparent conducting electrodes applications [6] as well as for energy storage [8]. However, despite its attractive properties, the lack of finite gap has been its main caveat in nanoelectronic applications [9]. It also exhibits severe aggregation and restacking which results in a much lower specific surface area. Low specific surface area leads to ions not accessing the surface of the electrode, and this affects an electrodes' cyclic ability [10]. Additionally, Gr has low storage capacity for alkali ions [11][12].

Two-dimensional transition metal dichalcogenides (2D TMDCs) on the other hand, are a family of materials whose generalized formula is $MX_2$, where M represents transition metal and X represents the chalcogenide elements. These materials are almost as thin, transparent and flexible as graphene, however unlike graphene, TMDCs have diversity of chemical compositions and structural phases that results in a broad range of electronic properties, both from the point of view of the emergence of correlated and topological phases and of the band structure character (metallic or insulating) [13]. Existence of semiconductor TMDCs means that they have the prospects for a wide range of applications [14]. $HfS_2$ is one such TMDC with an indirect energy band gap of ~1.252 eV [15], a good upper limit of mobility (~1800 cm$^2$/V$^{\cdot s}$) [16], and bonds that are more ionic than those in $MoS_2$ [17]. As a result, the charge transfer per S atom in $HfS_2$ is expected to be higher [17]. Another property of TMDC is that they posses weak van der Waals interaction between TMDCs layers, this make it possible to stack different TMDCs layers to form heterostructures with new electronic properties. Graphene based heterostructures have been created by using graphene as one of the layers forming the heterostructure. This has already been done in the case of $Gr/MoS_2$ [18], $Gr/WS_2$ [19] and $Gr/VS_2$ [20].

Studies have reported the possibility of alkali ions intercalation in these van der Waals heterostructures with binding energies per intercalated ion as well as band gap increasing with increase in the number of intercalated ions [21][22]. Alkali ion intercalation has also been found to lead to the vertex of the Dirac cone shifting downward due to *n*-doping of the Gr monolayer by the electrons transferred from intercalated atoms [20]. In addition, such heterostructures have the potential to overcome the restacking problem of pure Gr [23].

In this study, using dispersion corrected density functional theory (vdW-DFT), alkali ion intercalation in $Gr-HfS_2$ van der Waals heterostructure has been investigated to determine; the interlayer binding energy, identify the minimum energy configuration of the $Gr-HfS_2$ heterostructure as well as investigate the influence of intercalants (Li, Na and K) on the properties of the $Gr-HfS_2$ heterostructure, among others.

**2.0 Computational details**

In this work, first-principles calculations were performed within the density functional theory (DFT) framework, as implemented in Quantum ESPRESSO code [24]. The study used the Perdew–Burke–Ernzerhof (PBE) functional [25] to describe the electrons exchange-correlation potentials. Interlayer van der Waals (vdW) interactions of the $Gr-HfS_2$ systems were considered in all the calculations through the Van der Waals density functional (vdW-DF2) scheme [26]. To include the electron-ion interaction, norm-conserving pseudopotentials [27] were used for all the atoms. Monolayers of Gr and $HfS_2$ were obtained from their bulk counterparts whose equilibrium properties were obtained using a converged kinetic energy cut-off of 70 Ry, Gamma-centred k-point mesh of 8×8×3 for graphite and 7×7×4 for $HfS_2$. A convergence criteria of $10^{-6}$ Ry in calculated total energies was imposed on all the systems investigated. The optimized lattice constants were found to be 2.46 Å for Gr and 3.64 Å for $HfS_2$. The optimized c value was also obtained as 5.82 Å for $HfS_2$ and 6.71 Å for Gr. These values were in good agreement with previous studies which reported lattice constants of 2.47 Å for Gr [28], and 3.64 Å for $HfS_2$ [29].

The monolayer unit cells of Gr and $HfS_2$ were then created from the bulk systems and a 15 Å vacuum was added along the direction perpendicular to the atomic planes of the bulk structures of graphite and $HfS_2$, respectively. The vacuum helps to minimize the interaction between the layers along the *c*-axis. The atomic positions of the monolayer systems were relaxed keeping the volume fixed. The heterostructure was then constructed by placing the Gr monolayer on top of the $HfS_2$ monolayer. However, due to the difference in the equilibrium lattice constants of Gr and $HfS_2$, there was need to reduce the lattice mismatch in the created heterostructure. This was done by creating supercells of different sizes for each of the monolayers. Supercell sizes of 3×3×1 and 2×2×1 for Gr and $HfS_2$, respectively, were used in creating the heterostructure as this is what resulted in a small lattice mismatch of 1.37% between the Gr and $HfS_2$ layers. First–principles calculations with the climbing image nudged elastic band (CI–NEB) [30] method, as implemented in the Quantum ESPRESSO transition state tools was employed to investigate the energy barrier associated with the migration of the Li, Na and K atoms through the heterostructure. For comparison, diffusion through bilayer Gr and bilayer $HfS_2$ was also considered.

## 3.0 Results and discussion

Different orientations of Gr layer on top of $HfS_2$ layers (hereto referred to as configurations) were considered to determine the best Gr-$HfS_2$ configurations. To this end the heterostructure binding energy ($E_b$) was used as a descriptor, $E_b$ was defined as

$$E_b = \frac{E_{\text{Gr-HfS}_2} - (E_{\text{Gr}} - E_{\text{HfS}_2})}{N_C} \qquad 1.0$$

where, $E_{\text{Gr-HfS}_2}$, $E_{Gr}$ and $E_{\text{HfS}_2}$ are the calculated total energies of the Gr-$HfS_2$ heterostructure, Gr monolayer and $HfS_2$ monolayer, respectively, and $N_c$ is the total number of C atoms in the system. By this definition (equation 1.0), the configuration with the lowest binding energy was selected. Among the considered configurations, included having the system using the lattice parameters of Gr as the reference, $HfS_2$ as a reference as well as the average of the lattice parameters of Gr and $HfS_2$ as reference. As seen in Table 1, the system with Gr as a reference lattice parameter had the lowest binding energy and was thus selected for subsequent calculations. In this case the $HfS_2$ layer was strained by about 1.37%.

Table 1: Binding energies corresponding to various Gr-$HfS_2$ heterostructure configurations. $E_b$ is the binding energy per Carbon atom

|  | Gr as the reference | $HfS_2$ as a reference | Gr and $HfS_2$ as reference |
| --- | --- | --- | --- |
| $E_b$ | - 0.040 eV | 0.038 eV | - 0.017 eV |

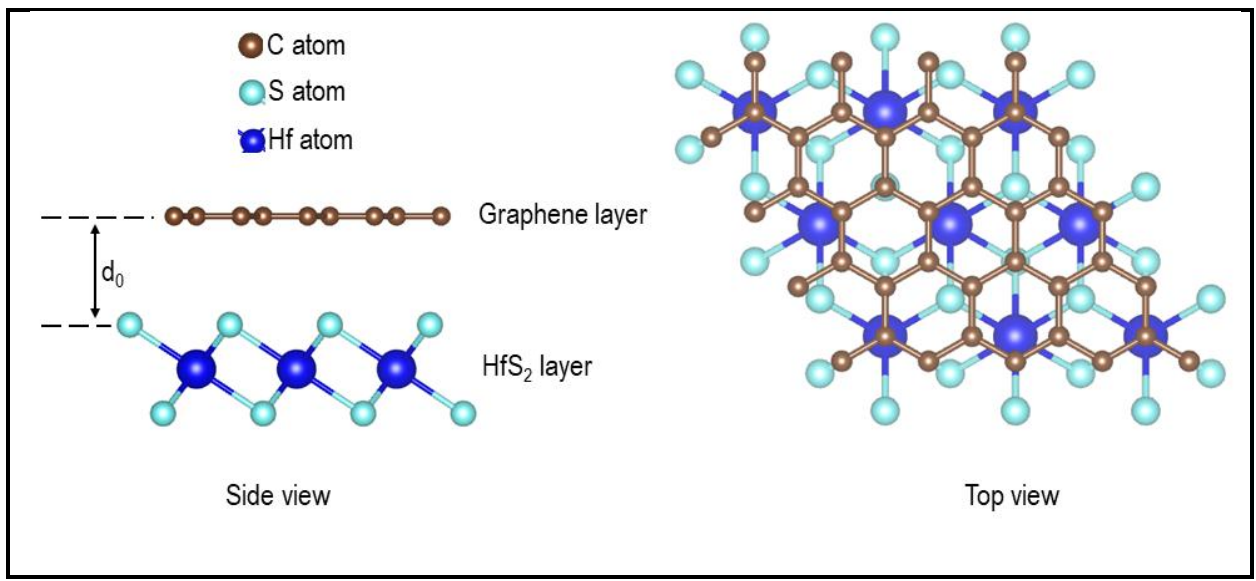

Figure 1: Schematic illustration of the Gr-$HfS_2$ heterostructure with the most energetically stable configuration.

The binding energy per C atom ($E_b$) was then used as a descriptor to obtain the equilibrium interlayer distance $d_0$ between the layers forming the heterostructure, (see, Figure 1). This was done by calculating the $E_b$ using equation 1.0 at different vertical distances, $d_0$. By this definition (equation 1.0), a lower $E_b$ value means a more stable heterostructure and vice versa. The calculated value of binding energy per C atom, $E_b$, at different interlayer distances are presented in Figure 2.

Figure 2 is a Lenard Jones type [31] of presentation and it indicates the presence of vdW interaction between the two layers of the Gr-$HfS_2$ heterostructure. The optimized equilibrium interlayer distance $d_0$ was found to be 3.00 Å and the corresponding binding energy was -140 meV. In other analogous systems $d_0$ was found to be 3.33 Å for bilayer Gr [32], 3.1 Å for $MoS_2$/Gr systems [32] and 3.22 Å for hexagonal-Boron Nitride/Gr (h-BN/Gr) hetero-bilayer [33]. The negative binding energies confirm the thermodynamic stability of the heterostructure. All subsequent calculations, were done using the obtained $d_0$.

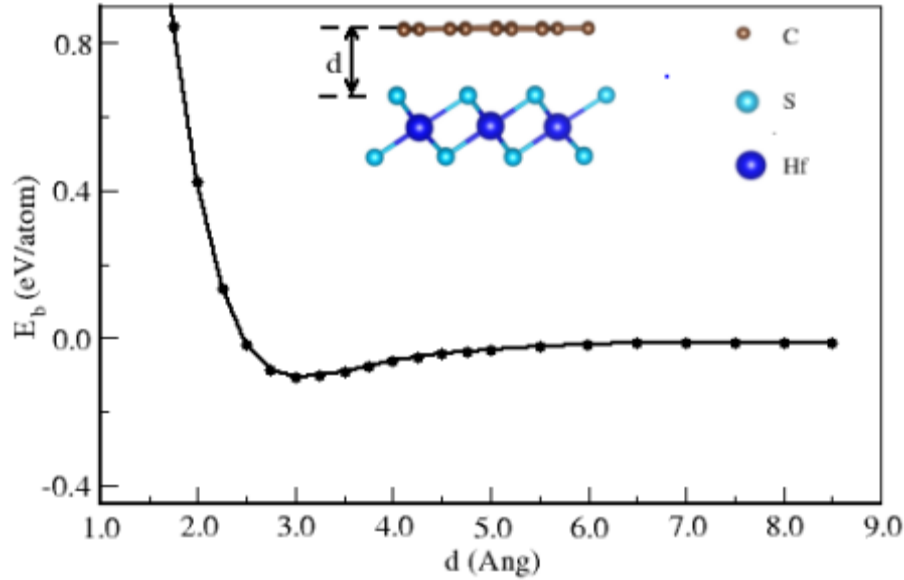

Figure 2: Binding energy of Gr-HfS$_2$ van der Waals heterostructure as a function of the interlayer distance d. Image inset shows the interlayer distance d.

## 3.1 Electronic properties

The calculated band structures and their respective DOS and PDOS for Gr, HfS$_2$ and Gr-HfS$_2$ heterostructure are shown in Figures 3, 4 and 5. Gr is semi metallic while HfS$_2$ has a wide band gap of 2.27 eV [34]. The monolayer of HfS$_2$, (Figure 3 (b)), was found to have a direct electronic band gap of 1.45 eV, which compares well with a previous study that found the band gap to be 1.28 eV [35]. As can be seen in Figure 3, the weak interaction between the two layers in the Gr-HfS$_2$ vdW heterostructure resulted in a vanishingly small bandgap (30.7 meV) opening at Gamma point. This observation is also consistent with previous graphene based heterostructures where electronic band gaps of the same order were observed. As examples, Pelotenia *et al* [33] observed an electronic band gap in hexagonal Boron Nitride/Gr hetero-bilayer of 20 meV, while Yuan *et al* [36] found a band gap of 11meV for Gr/WS$_2$. Other studies have also found equally small band gaps such as 0.4meV for Gr/MoS$_2$ [37].

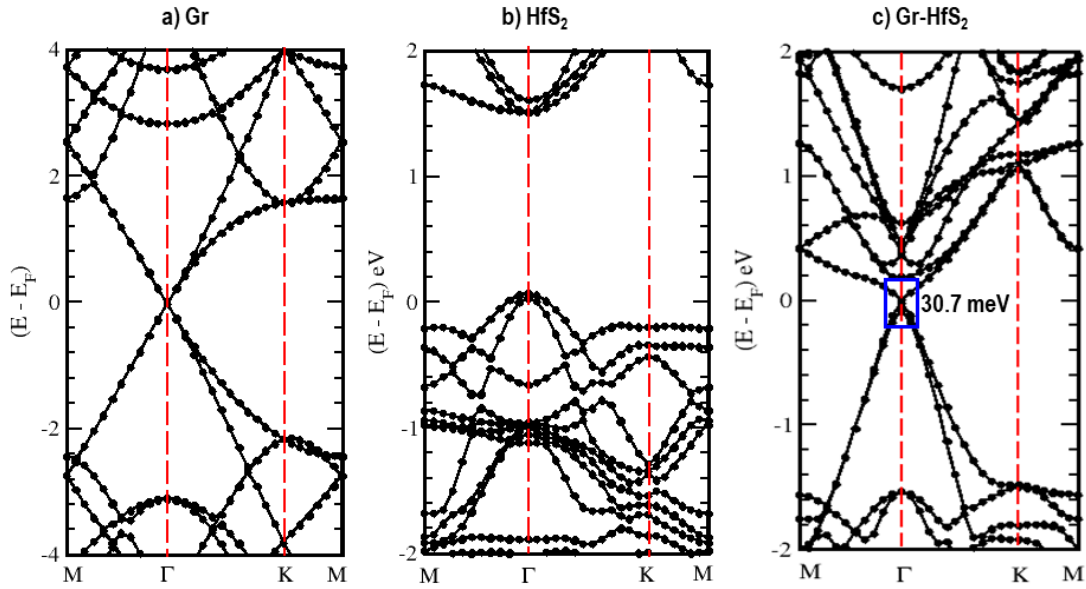

Figure 3: Calculated electronic band structures of 3×3×1 Gr supercell, 2×2×1 HfS$_2$ supercell and Gr-HfS$_2$ heterostructure system. Gr has no electronic band gap, HfS$_2$ has an electronic band gap of 1.45 eV while Gr-HfS$_2$ has an electronic band gap of 30.7 meV.

From the DOS (Figure 4) fewer states were occupied near the Fermi level in the case of Gr and Gr-HfS$_2$ heterostructure plots and this was consistent with the observations of Figure 3, where Gr was found not to have a band gap while Gr-HfS$_2$ had a band gap of 30.7 meV. The atomic orbital contributions on the band edges are shown in the PDOS plots (Figure 5). The *p* orbital of C in Gr was found to dominate the edges of the Dirac cone in graphene's band structure while the *d* orbital of Hf formed the conduction band edge of both HfS$_2$ monolayer as well as Gr-HfS$_2$ heterostructure. Electronic conductivity and ionic conductivity play a significant role during the intercalation/deintercalation of charge-carrying ions within an electrode material, since it influences the efficient movement of electrons and ions especially at high current rates [38]. The negligible electronic band gap of the Gr-HfS$_2$ heterostructure would therefore be expected to lead to efficient movement of electrons in the electrode.

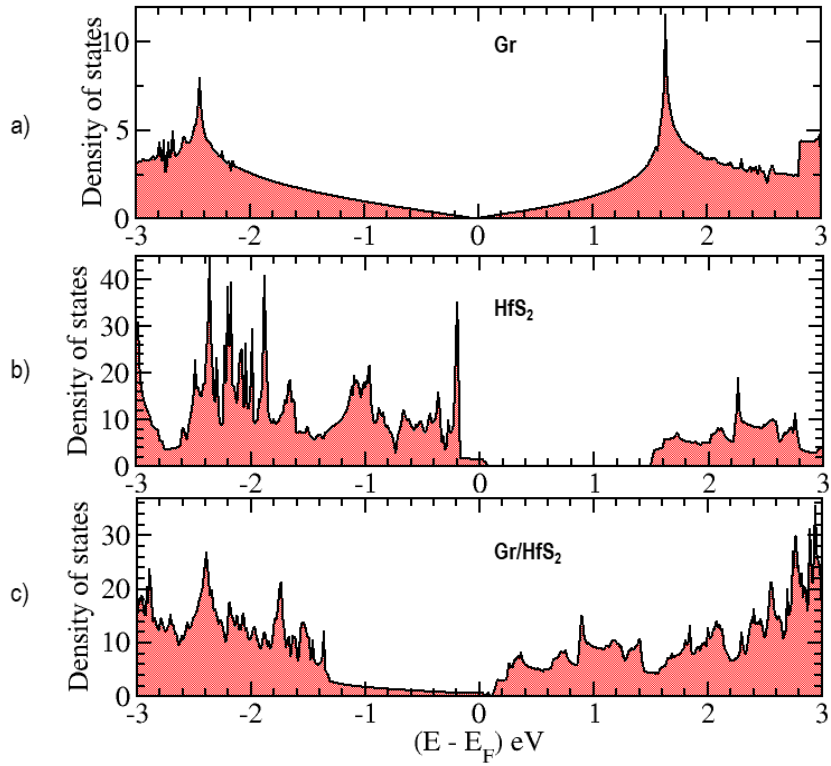

Figure 4: Calculated density of states of (a) 3×3×1 Gr supercell, (b) 2×2×1 HfS$_2$ supercell and (c) Gr-HfS$_2$ heterostructure system.

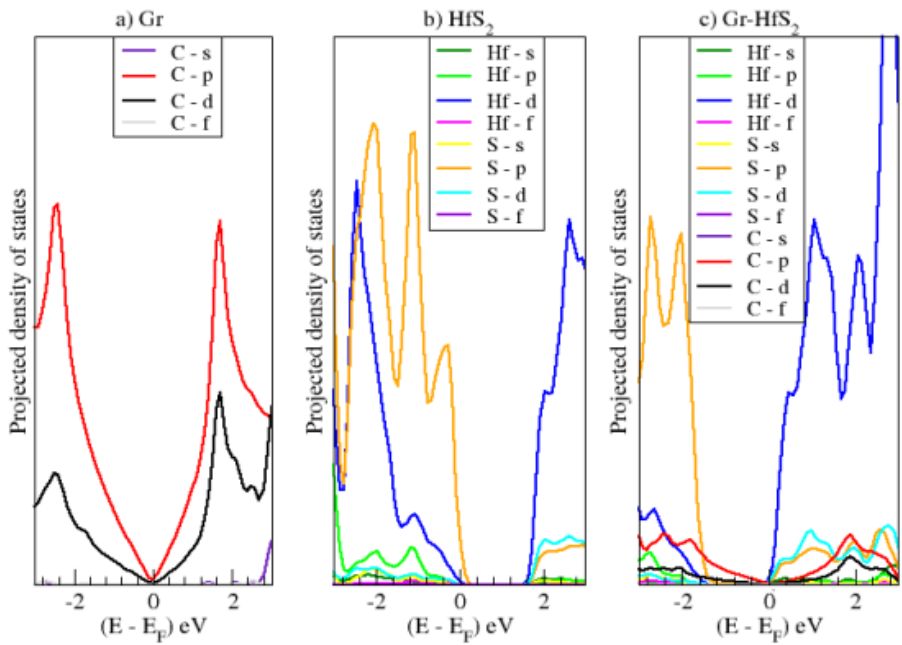

Figure 5: Calculated projected density of states of (a) 3×3×1 Gr supercell, (b) 2×2×1 HfS$_2$ supercell and (c) Gr-HfS$_2$ heterostructure system.

## 3.2 Work-function of the heterostructure

The electrostatic potential of the Gr-HfS$_2$ heterostructure was obtained along the z-direction (Figure 6). The vacuum level is the region outside the surface where the potential reaches a constant (flat level) and this was determined from the calculated macroscopic and planar averages of the electrostatic potential. The work function was calculated using the equation,

$$\Phi = E_{vac} - E_F \qquad\qquad 2.0$$

where $E_{vac}$ is the electrostatic potential in the vacuum region while $E_F$ refers to the Fermi energy [39].

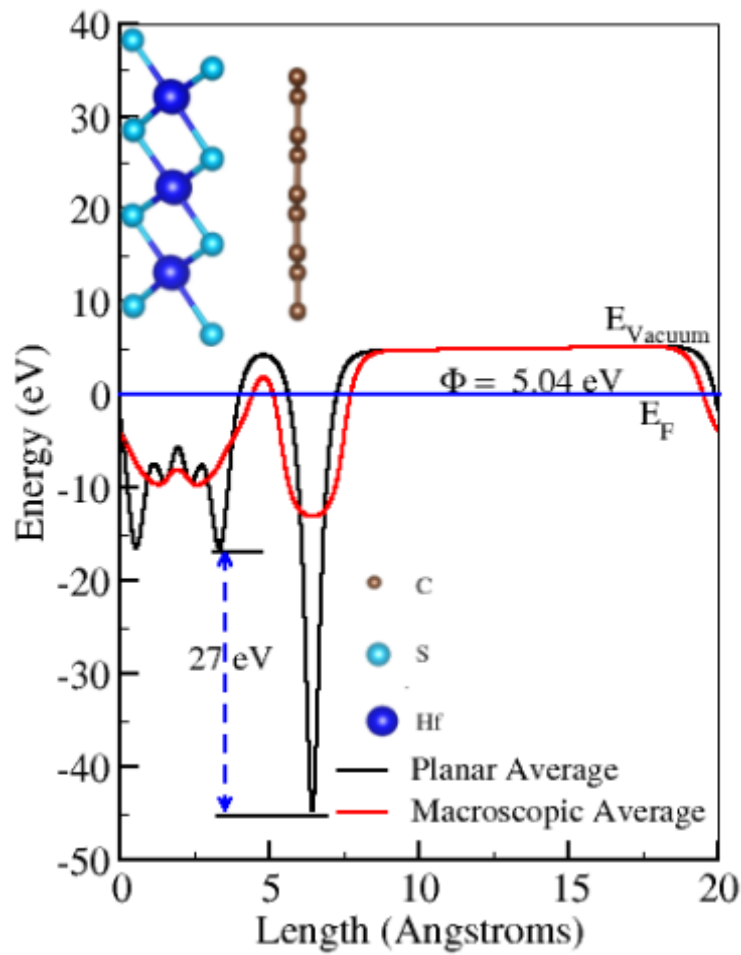

Figure 6: Planar and macroscopic electrostatic potentials for Gr-HfS$_2$ heterostructure, work function of the heterostructure is 5.04 eV. The image inset illustrates the atoms corresponding to various sections of the potential, the region between 6 Å and 20 Å is the vacuum region of the Gr-HfS$_2$ heterostructure.

The calculated values of the work function for Gr and $HfS_2$ were 4.25 eV and 6.20 eV, respectively, which were both in good agreement with previous studies [40]. The calculated work function for the Gr-$HfS_2$ heterostructure was 5.04 eV, implying that Gr decreases the work function of $HfS_2$ upon formation of the heterostructure, this in turn, makes it easier for electrons to be lost to the surface. The planar average potential around Gr consisted of a single distinct hump that corresponded to the monolayer of Gr, while the part around the $HfS_2$ consisted of three (3) peaks corresponding to the three sublayers of S, Hf and S, respectively. The electrostatic potential of Gr is deeper compared to that of $HfS_2$, and this results in a large potential drop of 27 eV across the *z*-direction of the heterostructure. This can be attributed to the differences in the atomic electronegativity of S = 2.58, Hf = 1.3 and C = 2.5. Hence, it is expected that electrons will be transferred from the Gr layer to the $HfS_2$ layer [41]. The large potential drop of 27 eV suggests a powerful electrostatic field across the interface, so that when the Gr layer is used as an electrode, this field will considerably affect the carrier dynamics and induce a low charge-injection barrier which will facilitate charge injection [28].

### 3.3 Alkali ion intercalation

Intercalation is the reversible insertion process of foreign species into the gap/space of a crystal. Layered materials are good host materials for various intercalant species ranging from small ions, to atoms and even to molecules [42]. Layered crystals are particularly suitable for intercalation processes as they can strongly adsorb guest species into their van der Waals interlayer spacing(s) [42]. In this study, the alkali ion(s) were inserted between the two layers of the Gr-$HfS_2$ heterostructure. A systematic study of intercalating different alkali ion species namely Li, Na and K in the Gr-$HfS_2$ heterostructure was carried out. This was informed by the fact that alkali ions such as Li have low reduction potentials that make their intercalation in battery materials attractive. Li is also the third lightest element with one of the smallest ionic radius of 2.20 Å [43]. The ionic radii of the other two alkali atoms, Na and K, are 2.25 Å and 2.34 Å, respectively [43]. It was anticipated that these other alkali ions, that is Na and K, might have similar properties as Li and hence the reason for their inclusion in this study. In addition and more importantly they are considerably more accessible than lithium [44].

The most energetically favorable position for the intercalants (with Li used as a test case) was established through the calculation of the binding energy with the intercalant in different positions. The binding energy was calculated as,

$$E_b = \frac{(E_{\text{GHfS}_2-\text{nM}} - E_{\text{GHfS}_2} - nE_M)}{n} \qquad 3.0$$

where $E_{\text{GHfS}_2-\text{nM}}$ is the total energy of the Gr-HfS$_2$ heterostructure with the alkali adatom, $E_{\text{GHfS}_2}$ is the total energy of the Gr-HfS$_2$ heterostructure without any alkali adatom, $E_M$ is the total energy of the free metal adatom, and *n* corresponds to the number of alkali ions.

The binding energy for the system when Li is adsorbed on Gr, (see figure 7(a)), above HfS$_2$, (see figure 7(b)), and when intercalated between the Gr-HfS$_2$ heterostructure layers, (see figure 7(c)), was found to be 0.4 eV, -1 eV and -1.6 eV, respectively, indicating that the system with Li between the layers is most stable.

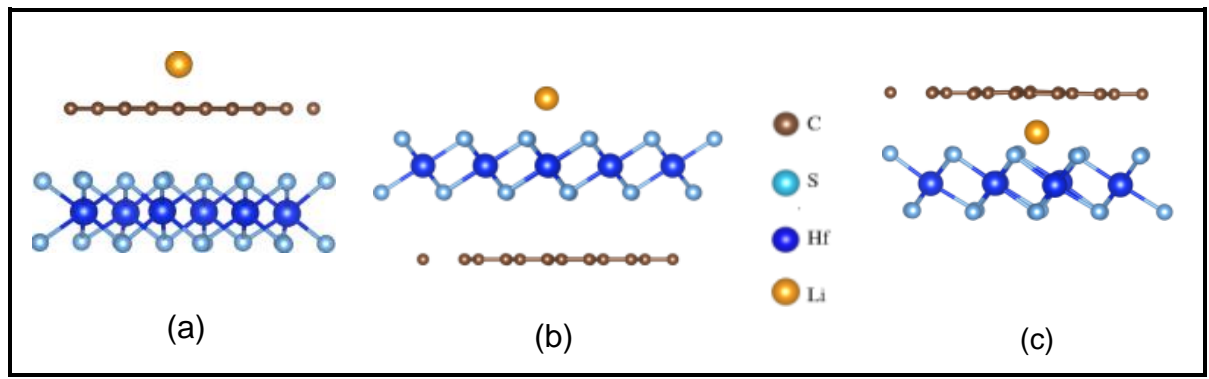

Figure 7: Side views of adsorption of Li atom on the Gr-HfS$_2$ vdW heterostructure, (a) Li above Gr (b) Li above HfS$_2$ (c) Li between Gr and HfS$_2$ layer.

Intercalation of alkali ions into the heterostructure was therefore done using the identified most stable Li configuration, that is, between the Gr-HfS$_2$ heterostructure layers (see Figure 7(c)). The preferred intercalation site(s) was then identified by intercalating a single Li adatom on the various available sites corresponding to the HfS$_2$ layer atoms at positions A, B, C, D, E as shown in figure 8 (a). Other sites used were the bridge position between two carbons, within the hollow site formed by the ring of Graphene carbon atoms and on the top position of Carbon atoms, (positions E, F and G, respectively) as shown in figure 8 (b), the site with the minimum binding energy, in this case A, as seen from Table 2 was considered the most stable and hence the one used as the intercalation site for the single intercalant in our calculations. It is worth noting however that there was negligible difference in the binding energy of the various adsorption sites.

During intercalation of two atoms, we identified two configurations that could be used to intercalate the atoms, that is either at positions A and B or points A and C, as illustrated in figure 8 (c). The binding energy associated with positions A and B was - 0.172 eV while that obtained for positions A and C was - 0.171 eV. As a result, intercalation of two atoms was done using a configuration similar to that of positions A and B. Only one configuration was possible for the 3 and 4 atoms (see figure 8 (c) and 8 (d)). The number of intercalated ions was progressively increased from 1 to 4 as there were no other equivalent sites available within the constructed Gr-HfS$_2$ heterostructure.

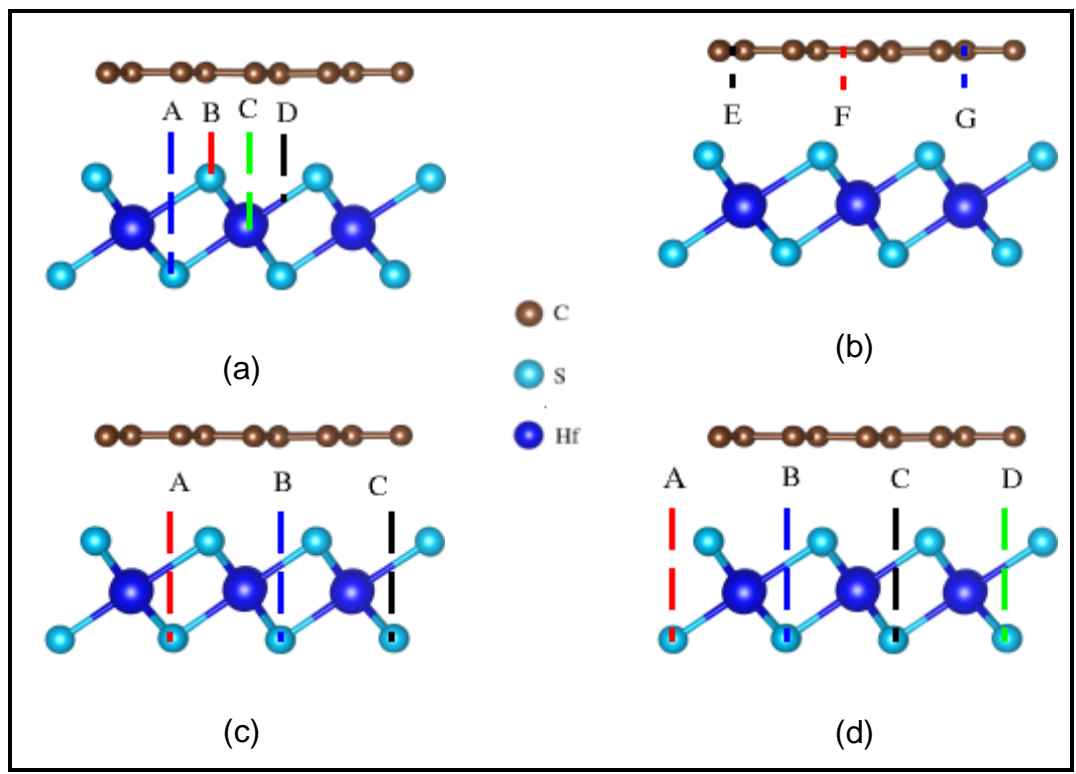

Figure 8: (a) and (b) Intercalation sites used to identify minimum energy positions (c) Intercalation sites used to identify minimum energy combination of two atoms. Intercalation of three atoms uses all the positions A, B and C as the intercalation sites. (d) Intercalation sites for four atoms used all the sites indicated as A, B, C and D.

Table 2: Binding energies corresponding to various adsorption sites in the Gr-HfS$_2$ heterostructure. E$_b$ is the binding energy (eV) per C atom

| position | A | B | C | D | E | F | G |
|---|---|---|---|---|---|---|---|
| E$_b$ | -0.091 | -0.089 | -0.088 | -0.079 | -0.090 | -0.088 | -0.089 |

### 3.4 Effect of intercalant concentration

The intercalation of alkali atoms in the Gr-HfS$_2$ heterostructure had an influence on the workfunction of the heterostructure, and this is a desirable property in energy storage media. As seen in Figure 9, the workfunction, calculated using equation 2.0, dropped with increasing alkali ion intercalant concentration up to a constant value of 4.58 eV for both the Li and K intercalant species, and 4.59 eV for Na intercalant.

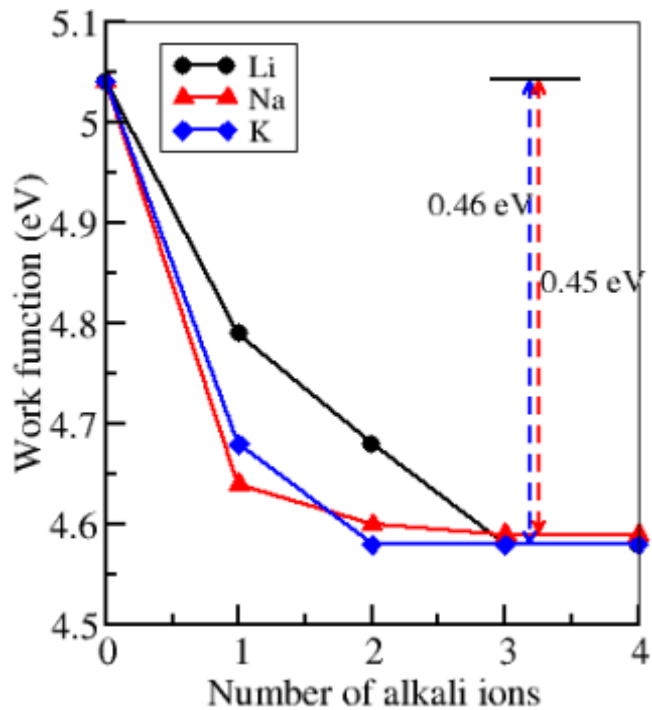

Figure 9: The workfunction of the Gr-HfS$_2$ heterostructure intercalated with the adatoms of Li, Na and K.

Upon reaching this constant value, the workfunction of the heterostructure had reduced by 460 meV in the case of Li and K intercalation and 450 meV for Na intercalation. This observation is consistent with other studies including Kim *et. al* [45], who observed that hole doping in Gr leads to a difference in the workfunction by as much as 400 meV. When the workfunction attains a constant value it is an indication that there is no more charge imbalance in the system resulting in no further electron flow.

The study also considered how the binding energy and interlayer distance varied as a function of the number of intercalated atoms (See Figure 10). The binding energy of the intercalated systems is observed to be highest for Li intercalated system and lowest for Na intercalated system, however, the binding energy values for all the concentrations for the Li, Na and K intercalated Gr-$HfS_2$ heterostructure are always negative, (see figure 10 (a)). This suggests that Li, Na and K intercalation in the Gr-$HfS_2$ heterostructure is indeed stable, and no phase separation into individual monolayers or the formation of bulk alkali metals is expected.

The binding energies per alkali adatom, (see figure 10 (b)), gradually reduce with increasing concentration of the intercalated alkali intercalants. This is in line with the behaviour observed in Figure 9, where an increase in the number of intercalated adatoms results in reduction of the workfunction. The reduction in binding energy per alkali adatom, can be attributed to weak electrostatic attraction between the Gr-$HfS_2$ host and the alkali adatoms as a result of enhanced alkali-alkali repulsion as the concentration of intercalants is increased. As the number of adatoms is increased the inter-atomic distances between positively charged atoms reduces. For the Li atom the binding energy per Li atom decreases from -1.6 eV to -1.4 eV as the number of intercalated atoms increases from 1 to 4. This can be attributed to the enhanced repulsive interaction between the positively charged Li ions. For K intercalation, the binding energy per K atom initially increases from -0.9 eV to -1.3 eV upon introduction of the first and second K atoms and then decreases. This observation is consistent with an observation made by Demiroglu *et. al* [46] for K intercalation in $Ti_2CO_2$ Mxene/Gr heterostructure [46]. For K and Na adatoms intercalation, the binding energy per K/Na atoms is initially very low as compared to that of Li. This can be attributed to the fact that the larger size of K and Na ions distorts the lattice of the Gr-$HfS_2$ heterostructure in comparison to Li ions.

The change in the interlayer distance between the two layers forming the Gr-$HfS_2$ heterostructure increases with increasing number of Li ions peaking at 3 Li ions and decreases at 4 Li ions intercalation. For Na and K ions intercalations, the peak was at two ion intercalations, (see Figure 10 (c)). Additionally, the maximum increase in the interlayer separation was found to be 0.21 Å for Li, 0.94 Å for Na and 1.7 Å for K which corresponded to volumetric expansion in the z-direction in the order of 6%, 31% and 56.3%, respectively. The 6% volumetric expansion in the case of Li intercalation is comparable with that of graphite anodes which is 10% [47]. The 31% and 56.3%, for Na and K atoms intercalation is much lower than that for silicon based electrodes which is 280% [48] or for alloy-type anodes which is 260% for Germanium (Ge) and Tin (Sn), and 300% for Phosphorus (P) [49]. These observations indicate that the Gr-HfS$_2$ heterostructure is likely to possess a reversible reaction process in the case of Li, Na and K intercalation, which is an essential property for rechargeable ion batteries. This attribute also implies that Li intercalation in Gr-HfS$_2$ heterostructure effectively overcomes the volume expansion problem faced by electrode materials.

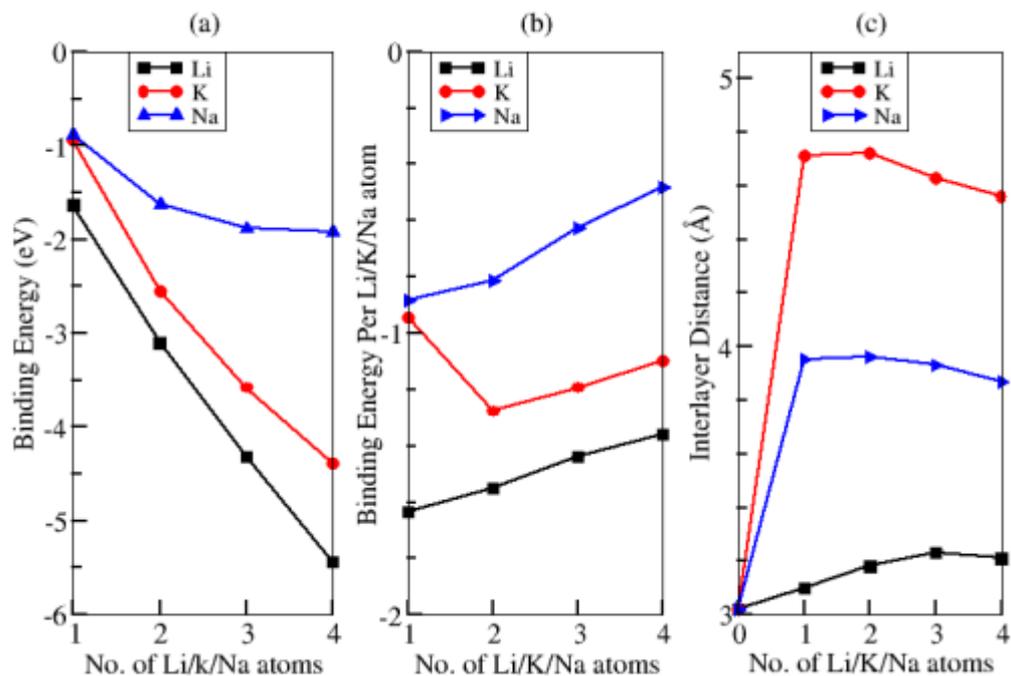

Figure 10: (a) Binding energy as a function of K, Na and Li concentration in the intercalated Gr-HfS$_2$ heterostructure (b) Binding energy per Li/K/Na atom as a function of K, Na and Li concentration in the intercalated Gr-HfS$_2$ heterostructure and (c) interlayer distance as a function of K, Na and Li concentration in the intercalated Gr-HfS$_2$ heterostructure

## 3.5 Alkali atom diffusion through the heterostructure

The charge/discharge rates of metal-ion batteries predominantly depend on the ion diffusion in the electrode materials, which further determines the mobility of the adatoms, since a smaller energy barrier would facilitate faster diffusion. Poor diffusivity leads to significant structural damage with continued cycling, consequently affecting the lifetime of the battery [50].

To investigate the migration of the Li, Na and K atoms through the heterostructure, we first located the lowest energy site and then studied the pathways between this site and adjacent sites. Based on the length of the pathways, 3 - 5 images were employed between various distinct paths as shown in figure 11. The minimum energy path between the two adjacent points gave the energy barrier between them.

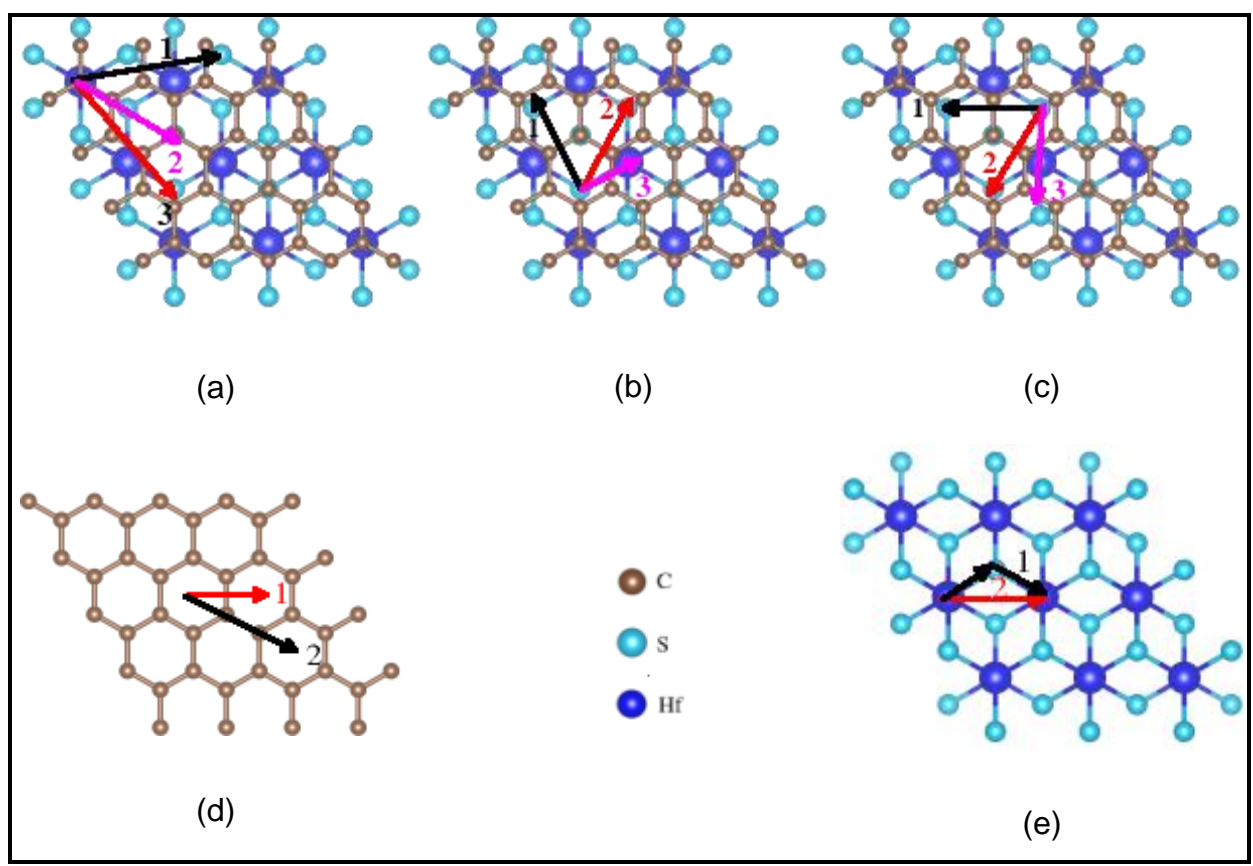

Figure 11: (a) Migration pathways of the Li adatom in the Gr-HfS$_2$ heterostructure (b) Migration pathways of the Na adatom in the Gr-HfS$_2$ heterostructure (c) Migration pathways of the K adatom in the Gr-HfS$_2$ heterostructure (d) Migration pathways of the Li, Na and K adatom in the Gr bilayer (e) Migration pathways of the Li, Na and K adatom in the HfS$_2$ bilayer. The arrows 1, 2 and 3 show the sites between which the alkali adatoms move in each case. Arrow directions indicate the direction of the movement of the adatom

The energy barriers associated with the intercalants during their migration using different paths are presented in Table S1 (In the supplementary information). It was found that for the Li intercalated heterostructure the energy barrier varies between 0.22 eV and 0.39 eV, between 0.05 eV and 0.09 eV for K intercalated heterostructure and between 0.32 eV and 0.50 eV for Na intercalated heterostructure. The diffusion of potassium through the Gr-$HfS_2$ heterostructure only varied negligibly through the various paths. Figure 12 shows the diffusion energy barriers for the considered systems when we consider the lowest energy barrier path for each of the systems. From Figure 12, we note that the diffusion energy barriers are lower in the Gr-$HfS_2$ heterostructure compared to both bilayer Gr and $HfS_2$ with the exception of Na diffusion. The diffusion energy barriers on the heterostructure system are lower for Na and K ions than Li for the respective minimum energy pathways due to the stronger binding of Li intercalation as seen in Figure 10. Strong binding energies are expected to pin the atoms on the surface. In order to move the intercalant between sites, a certain amount of energy is required to overcome the adsorption interaction at the site. Hence moving Li, which is the most strongly bonded metal, requires a larger energy threshold to be overcome than the equivalent process for Na and K. The increased interlayer distance in the potassium intercalated heterostructure is also expected to enhance the diffusion process, leading to the potassium intercalated heterostructure having the lowest energy barrier. From the values of Table 3, the minimum diffusion energy barrier associated with the intercalated heterostructure systems for Li, Na and K are, respectively, 0.22 ev, 0.28 eV and 0.05 eV, all these values are lower than for Li ion on graphite (0.42 eV) [51] and on commercially used anode materials based on $TiO_2$ (0.32−0.55) [52]. The lower diffusion energy barriers on the heterostructure systems indicates higher mobility and hence improved battery performance for the heterostructure.

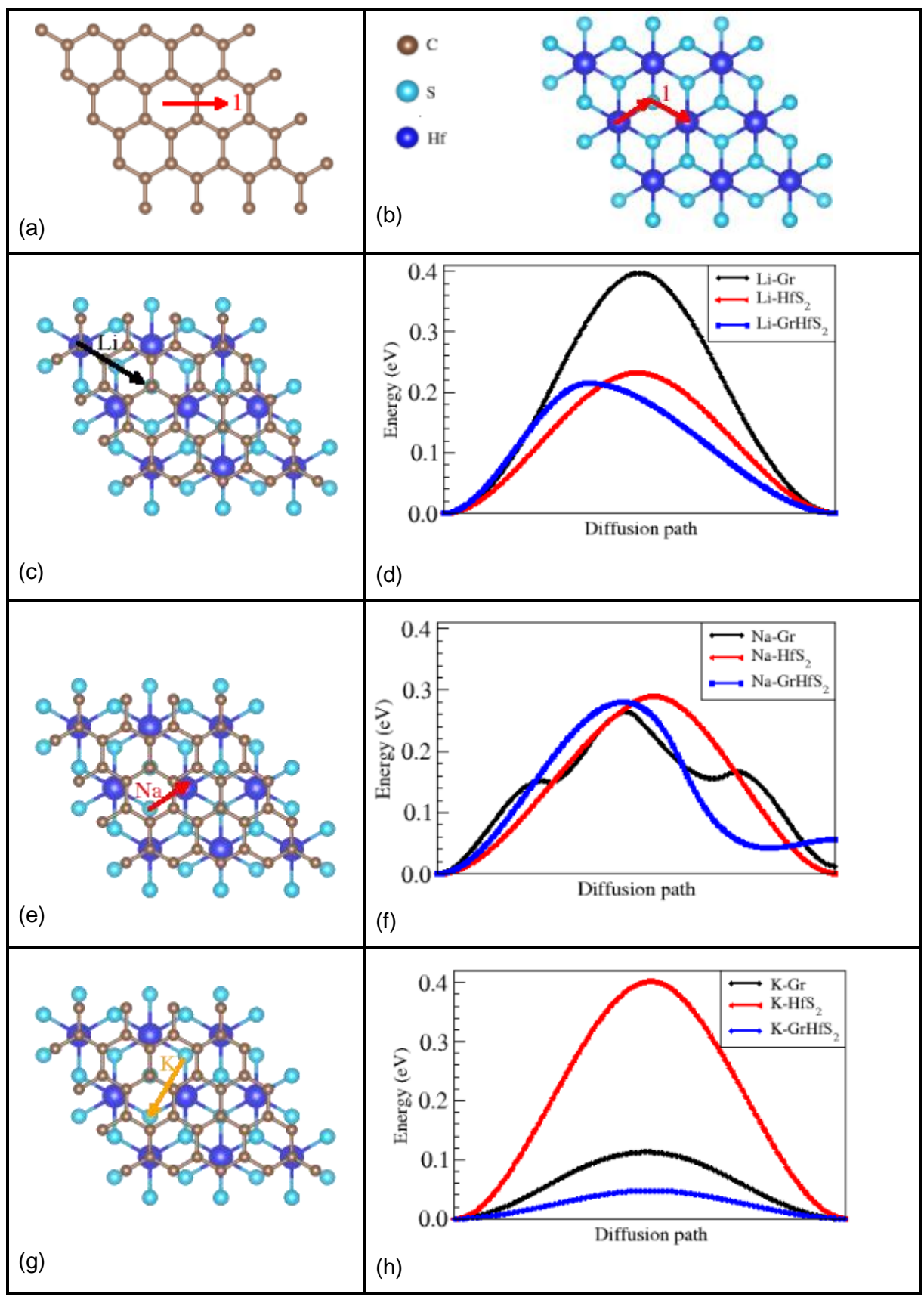

Figure 12: Minimum energy paths (MEP) and their associated energy profiles; (a) MEP through bilayer Gr for Li/Na/K intercalants, (b) MEP through bilayer $HfS_2$ for Li/Na/K

intercalants, (c) MEP through Gr-HfS$_2$ for Li intercalants, (d) Energy profile associated with Li diffusion through the paths shown in *a, b* and *c* for the bilayers of Gr, HfS$_2$ and Gr-HfS$_2$ heterostructure, respectively, (e) MEP through Gr-HfS$_2$ for Na intercalant, (f) Energy profile associated with Na diffusion through the paths shown in *a, b* and *c* for the bilayers of Gr, HfS$_2$ and Gr-HfS$_2$ heterostructure, respectively, (g) MEP through Gr-HfS$_2$ for K intercalant (h) Energy profile associated with K diffusion through the paths shown in *a, b* and *c* for the bilayers of Gr, HfS$_2$ and Gr-HfS$_2$ heterostructure, respectively. Only the lowest energy barrier profile is shown for the various systems.

### 3.6 Electrochemical properties

In order to gain insights into the electrochemical properties of the Li, Na and K intercalation process into the Gr-HfS$_2$ heterostructure, the open-circuit-voltage (OCV) was determined. The OCV value gives a measure of the performance of a battery, and was calculated from the energy difference based on the equation

$$V \approx \frac{\left[\left(E_{GrHfS_2+x_{1M}} - E_{GrHfS_2+x_{2M}}\right) + (x_2 - x_1)\mu_M\right]}{(x_2 - x_1)e} \qquad 3.0$$

where $E_{GrHfS_2+x_{1M}}$ and $E_{GrHfS_2+x_{2M}}$ are the total energies of the Gr-HfS$_2$ heterostructure with $x_1$ and $x_2$ alkali atom intercalated, respectively, $\mu_M$ is the chemical potential of Li/Na/K atom and *e* denotes the elementary charge quantity [53][54][55]. The chemical potential of Li/Na/K atom is approximately equal to the total energy per Li/Na/K atom, and hence this was the value used in equation 3.0 [53][55].

The calculated voltage profiles of the three considered systems are shown in Figure 13. It is observed that the voltage decreases gradually from 1.64 V to 1.36 V as the number of Li adatoms increases, while that of K intercalated system initially increases from 0.94 V to 1.28 V then decreases to 1.10 V. The calculated average voltage profile is 1.49 V for Li, 1.13 V for K and -2.66 V for Na intercalated systems. The voltage is positive for all Li and K concentrations, meaning that the Li and K intercalated system can be fully intercalated, the negative values for Na intercalated Gr-HfS$_2$ heterostructure indicate that Na intercalation is chemically unstable for the Gr-HfS$_2$ heterostructure. The calculated voltage values for all systems correlate with the binding energy values, presented in figure 10. The highest voltage is found for Li as this system has the largest binding energy, (see figure 10 (a)). The lowest voltage is found for Na as this system has the least binding energy, (see figure 10 (a)). Our results indicate that Li and K intercalation in Gr-HfS$_2$ heterostructure can be exploited in low voltage applications.

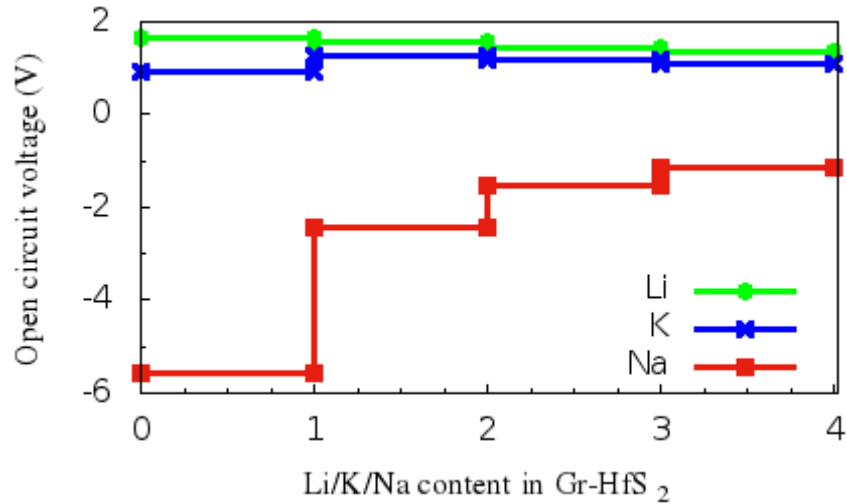

Figure 13. Open circuit voltage profiles of Li, Na, and K intercalation in Gr-HfS$_2$ heterostructures as a function of alkali atom concentration.

### 3.7 Charge density distribution

In order to understand the mechanism of the charge distribution and charge transfer between Gr and HfS$_2$ monolayers in the Gr-HfS$_2$ heterostructures, we calculated the charge densities difference (Δ $\rho$) using the relation:

$$\Delta\rho = \Delta\rho_{Gr/HfS_2} - \Delta\rho_{Gr} - \Delta\rho_{HfS_2} \qquad 4.0$$

where $\Delta\rho_{Gr/HfS_2}$ is the charge density of the heterostructure, $\Delta\rho_{Gr}$ is the charge density of Gr and $\Delta\rho_{HfS_2}$ is the charge density of Hafnium disulfide. The resulting charge density difference distribution is shown in Figure 14. Evidence of charge distribution between the two layers is observed with and without the Li intercalants. Only the Li interaction was considered in this step since it was the intercalant that resulted in the Gr-HfS$_2$ bilayer exhibiting desirable battery electrode properties. Charge accumulation is represented by the green iso-surface while charge depletion is represented by the red iso-surface. It is worth noting that the iso-surface level for the pristine Gr-HfS$_2$ heterostructure was 0.0004 e.Å$^{-3}$ while those of the rest was 0.4 e.Å$^{-3}$.

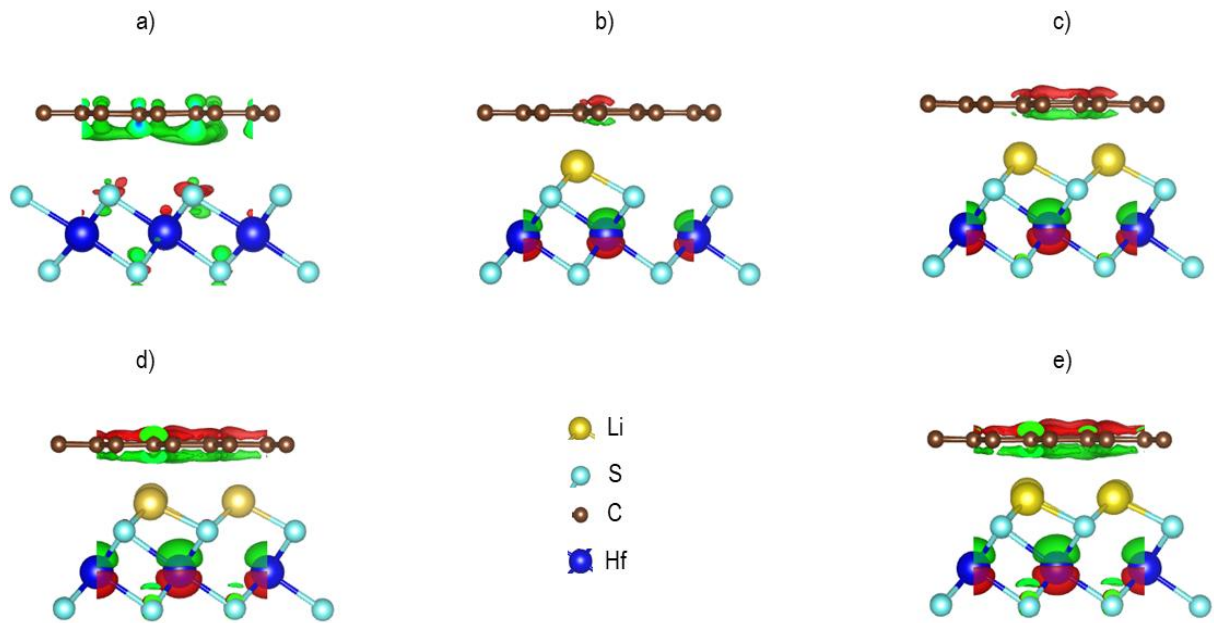

Figure 14: Charge density difference plot for the Gr-HfS$_2$ heterostructure systems (a) without Li intercalants (b) with ONE Li adatom intercalants, (c) with TWO Li adatom intercalants, (d) with THREE Li adatom intercalants, (e) with FOUR Li adatom intercalants. The green isosurface indicates charge accumulation while the red isosurface indicates charge depletion. The isosurface value is 0.004 e.Å$^{-3}$ for all the intercalated systems and 0.4 e.Å$^{-3}$ for the pristine system.

The iso-surface between Gr and HfS$_2$ layer is a charge accumulation region. As the number of intercalants increase, (as we move from Figure 14(b) to 14(d)) the amount of charge accumulated increases while the regions with depletion around Hf ions also increases, this can be attributed to the reduction in the workfunction of the heterostructure as the number of intercalants is increased from one to four, (see figure 9). A reduction in the work-function makes it easier for electrons to be lost to the surface, hence the increase in the charge depletion and charge accumulation regions as the number of ions increases. Within the HfS$_2$ layer the charge accumulation mainly occurs around the Sulphur atoms, an indication that these atoms gain negative charges. A similar observation has been made for the Gr/MoS$_2$ [56] and tungsten sulfide (Ws$_2$)/Gr [57], heterostructures. It has also been shown that Tungsten diselenide (WSe$_2$) is a weak acceptor of electrons upon contact with Gr, in a WSe$_2$/Gr heterostructure [58].

**4.0 Conclusion**

This study has systematically investigated the prospects of Gr-HfS$_2$ heterostructure, as an electrode material for alkali ion (Li, Na and K) batteries, using first-principles calculations with vdW-DF corrections. The stability of the heterostructure upon alkali ion intercalation is confirmed by the negative binding energy values for all the intercalated atoms and also by donation of a significant amount of charge to the host material. The volumetric expansion due to the intercalant species was found to be 6%, 31% and 56.3%, for Li, Na and K, respectively, suggesting that the Gr-HfS$_2$ heterostructure possess a reversible reaction ability. Diffusion energy barriers confirm the advantage of Gr-HfS$_2$ heterostructure over Graphene and HfS$_2$ bilayer systems. Relatively low diffusion energy barriers ranging between 0.22 - 0.39 eV for Li, 0.05 - 0.09 eV for K and 0.28 - 0.74 eV for Na were determined for the intercalated Gr-HfS$_2$ heterostructure. This implies high charge/discharge rate in battery applications. Li intercalation in Gr-HfS$_2$ is attractive for rechargeable ion battery applications as it overcomes the volume expansion problem faced by many electrode materials. The findings of this study suggest that it is possible to develop next-generation anode materials with ultrafast charging/discharging rates using Gr-TMDC heterostructure.

**Acknowledgements**

The authors wish to acknowledge the CHPC South Africa for availing the supercomputing facilities utilized in carrying out this work.

Appendix A

**Supplementary information**

Table S1: Energy barriers associated with diffusing adatoms through the various paths shown in figure 11.

|  | Path 1 | Path 2 | Path 3 |
|---|---|---|---|
| *Li diffusion through GrHfS$_2$* | 0.39 eV | 0.22 eV | 0.26 eV |
| Na *diffusion through GrHfS$_2$* | 0.46 eV | 0.74 eV | 0.28 eV |
| K *diffusion through GrHfS$_2$* | 0.06 eV | 0.05 eV | 0.09 eV |
| Li *diffusion through Gr bilayer* | 0.39 eV | 0.42 eV | |
| Na *diffusion through Gr bilayer* | 0.26 eV | 0.54 eV | |
| K *diffusion through Gr bilayer* | 0.11 eV | 0.13 eV | |
| Li *diffusion through HfS$_2$ bilayer* | 0.10 eV | 0.69 eV | |
| Na *diffusion through HfS$_2$ bilayer* | 0.32 eV | 0.50 eV | |
| K *diffusion through HfS$_2$ bilayer* | 0.19 eV | 0.40 eV | |